\documentclass[12pt]{article}
\usepackage{amssymb}
\usepackage{amsmath}
\usepackage{mathrsfs,dsfont}
\usepackage{mathtools}
\usepackage{amsfonts}
\usepackage{cite} 
\usepackage[english]{babel}
\usepackage{hyperref}
\usepackage{xcolor}
\renewcommand{\theequation}{\arabic{equation}}
\newcommand{\be}{\begin{equation}}
\newcommand{\ee}{\end{equation}}
\newcommand{\bea}{\begin{array}}
\newcommand{\ea}{\end{array}}
\newcommand{\beqa}{\begin{eqnarray}}
\newcommand{\eeqa}{\end{eqnarray}}
\newcommand{\bean}{\begin{eqnarray*}}
\newcommand{\eean}{\end{eqnarray*}}

\def\up#1{\leavevmode \raise.16ex\hbox{#1}}

\newcommand{\gapproxeq}{\lower
.7ex\hbox{$\;\stackrel{\textstyle >}{\sim}\;$}}
\newcommand{\lapproxeq}{\lower .7ex\hbox{$\;\stackrel
{\textstyle <}{\sim}\;$}}
\renewcommand{\theequation}{\thesection.\arabic{equation}}
\newcounter{appendice}
\newcommand{\appendice}
{
\setcounter{equation}{0}
\renewcommand{\theequation}{\Alph{appendice}.\arabic{equation}}
\addtocounter{appendice}{1}
{\Large{\bf Appendix \Alph{appendice}}}
}
\def\thebibliography#1{{\bf REFERENCES\markboth
{REFERENCES}{REFERENCES}}\list
{[\arabic{enumi}]}{\settowidth\labelwidth{[#1]}\leftmargin\labelwidth
\advance\leftmargin\labelsep
\usecounter{enumi}}
\def\newblock{\hskip .11em plus .33em minus -.07em}
\sloppy
\sfcode`\.=1000\relax}

\def\BI{{\rm 1\!l}}

\begin{document}

\centerline{\LARGE Gauging the Schwarzian Action}
\vskip 1cm

\centerline{ A. Pinzul${}^1$\footnote{apinzul@unb.br}, A. Stern$^{2}$\footnote{astern@ua.edu} and   Chuang Xu$^{2}\footnote{cxu24@crimson.ua.edu} $ }

\vskip 1.5 cm

\begin{center}
  {${}^1$ Universidade  de Bras\'{\i}lia, Instituto de F\'{\i}sica\\
70910-900, Bras\'{\i}lia, DF, Brasil\\
and\\
International Center of Physics\\
C.P. 04667, Bras\'{\i}lia, DF, Brazil
\\}
\end{center}
\begin{center}
  {${}^2$ Department of Physics, University of Alabama,\\ Tuscaloosa,
Alabama 35487, USA\\}
\end{center}

\vskip 0.5cm

\centerline{\it This article is dedicated to the memory of A.P. Balachandran,}
\centerline{\it our teacher and friend.  }

\vskip 0.5cm

\abstract{	In this work, we promote the global $SL(2,\mathbb{R})$ symmetry of the Schwarzian derivative to a local gauge symmetry. To achieve this, we develop a procedure that potentially can be generalized beyond the $SL(2,\mathbb{R})$ case to any symmetry where the fundamental fields transform nonlinearly: We first construct a composite field from the fundamental field and its derivative such that it transforms linearly under the group action. Then we write down its gauge-covariant extension and apply standard gauging techniques. Applying this to the fractional linear representation of $SL(2,\mathbb{R})$, we obtain the gauge-invariant analogue of the Schwarzian derivative as a bilinear invariant of covariant derivatives of the composite field.  		The  framework enables a simple construction of Noether charges associated with the original global symmetry. The gauge-invariant Schwarzian action introduces $SL(2,\mathbb{R})$ gauge potentials, allowing for locally invariant couplings to additional fields, such as fermions. While these potentials can be gauged away on topologically trivial domains, non-trivial topologies (e.g., $S^1$) lead to distinct topological sectors. We mention that in the context of two-dimensional gravity, these sectors could correspond to previously discussed  defects in the bulk theory. 	}

\section{Introduction}

The  Schwarzian derivative\cite{Ovsienko,Osgood1998}  has  played a role  in many areas of theoretical physics.   For example, it controls the quantum corrections in the semi-classical approximation of quantum systems.
In  recent decades it has  appeared in  Virasoro algebras\cite{Witten:1987ty,Alekseev:1988ce,Alekseev:2022qrq,Chirco:2024qpx}, black hole physics\cite{Carlip:2022fwh,Lin:2019kpf}, the Sachdev-Ye-Kitaev (SYK) model\cite{PhysRevLett.70.3339,A. Kitaev,Maldacena:2016hyu}
 and Jackiw-Teitelboim (JT)  gravity\cite{T1983,J1985}. Moreover, the Schwarzian derivative  plays the central role in the proposed correspondence between the SYK model and  JT gravity.\cite{Maldacena:2016upp,Turiaci:2024cad}   It has been argued  to  result from  the SYK model   in the IR limit  and to represent the boundary dynamics of JT gravity.

The utility of the  Schwarzian derivative  is due in large part to its invariance under the  action of  $SL(2,\mathbb{R})$ (or more precisely, $PSL(2,\mathbb{R})\,$).  The group action  is global.  In this article we promote the $SL(2,\mathbb{R})$ target space symmetry to a gauge symmetry and write down the gauge invariant version of the Schwarzian derivative.\footnote{Certain gauge symmetry extensions of the base (rather than target) manifold  have been considered previously, and their relevance for a possible one-dimensional quantum gravity theory has been discussed.\cite{Bouarroudj,Anninos:2021ydw}} Aside from the intrinsic interest in gauging the Schwarzian derivative, it is of relevance in the SYK/JT correspondence. In the BF formulation of JT gravity, the bulk theory is described in terms gauge fields valued in the $sl(2,\mathbb{R})$ Lie algebra. Since the  Schwarzian derivative is associated with the boundary dynamics in JT gravity,   the resulting symmetry at the boundary is restricted to (global) $SL(2,\mathbb{R})$, in contrast  to the local symmetry of the bulk. The gauging of the Schwarzian derivative allows for  the $SL(2,\mathbb{R})$ symmetry  to be implemented locally on the boundary, extending the gauge redundancy and the gauge fields in the bulk to the boundary.\footnote{Other extensions of the boundary dynamics for JT gravity have been considered.\cite{Ozer:2025bpb}}   On topologically trivial domains, the boundary gauge field can be completely gauged away, and the theory reduces to the usual Schwarzian description. However, when the boundary is $S^1$ the local formulation naturally exposes additional topological sectors labeled by the holonomy of the $SL(2,\mathbb{R})$  connection, which are not present in  the original formulation.   From this perspective, the gauged Schwarzian does not modify the physical content of the boundary theory in the trivial sector.  Rather, it provides a natural framework for admitting the nontrivial  topological sectors at the boundary.
Upon extending these sectors to the bulk the flatness requirements of the interior are violated. This indicates that the curvature is locally sourced.  Such sources are generally associated with defects,  which are  of current interest.\cite{Mertens:2019tcm,Lin:2023wac,Turiaci:2024cad}
The gauging of the $SL(2,\mathbb{R})$ symmetry could also  be of interest on the SYK side of the correspondence. (A step in this direction was made in \cite{Zhang:2025kty} with the gauging of $U(1)$. Their approach  is to gauge internal symmetries in the complex SYK model in contrast to our approach of gauging  the geometric $SL(2,\mathbb{R})$ symmetry of boundary reparametrizations.)

The gauging procedure required for the Schwarzian derivative is nontrivial.  This is because the fundamental field $f$ that spans the domain of the Schwarzian derivative belongs to a nonlinear representation of the $SL(2,\mathbb{R})$  group, more specifically the fractional linear representation.  This complication can be overcome by constructing a composite field ${\bf f}$ from the fundamental field and its derivative  that then transforms linearly, more specifically, ${\bf f}$ belongs to the adjoint representation of $SL(2,\mathbb{R})$. The composite field has a number of useful features. The Schwarzian derivative ends up being a bilinear invariant of derivatives of ${\bf f}$. Noether charges for dynamical systems based on the Schwarzian derivative can be constructed from the composite field and its derivatives.   These Noether charges  are associated with the global $ SL(2,\mathbb{R})$ symmetry of the action ${\cal I}_0$ obtained from the Schwarzian derivative.

The gauging of the group is done by constructing the gauge covariant extension of ${\bf f}$, along with introducing its covariant derivative.   As usual, the covariant derivative  requires the introduction of  gauge potentials ${\bf A}$, here valued in the $sl(2,\mathbb{R})$ Lie algebra. The gauge invariant Schwarzian derivative is a bilinear invariant of  covariant derivatives of the covariant extension of ${\bf f}$.  Upon expanding the gauge invariant Schwarzian derivative to leading order in the gauge fields, one recovers the coupling to the Noether charges.
An action ${\cal I}[A]$ can be constructed from the gauge invariant  Schwarzian derivative which is the natural extension of ${\cal I}_0$.
  In the absence of any kinetic energy term for ${\bf A}$ in the Schwarzian action, the potentials can be completely gauged away when the domain for ${\bf A}$ is topologically trivial.  However, as discussed above, disconnected sectors arise when the domain is $\mathbb{S}^1$, leading to defects in JT gravity.  The focus of this article is primarily on the classical aspects of the gauge invariant Schwarzian, with quantum aspects left for future considerations.

The outline for the rest of this article is the following:
In section two we review the fractional linear representation of $SL(2,\mathbb{R})$, and give the  construction  of the composite field,  along with some of its properties.
 Dynamical systems  based on the Schwarzian derivative are examined in section three. There we construct  Noether charges (associated with the global $ SL(2,\mathbb{R})$ symmetry) from the composite field, as well as propose various $SL(2,\mathbb{R})$ invariant couplings to  external fields.  The gauge invariant Schwarzian derivative is presented in section four.  We examine  the possibility of replacing the boundary Schwarzian dynamics of JT gravity by its gauge invariant analogue in section five. (More specifically, we examine the gauge theory formulation of JT gravity, as opposed to the geometrical approach.  The two formulations are known to differ at the quantum level.\cite{Collier:2023fwi}) There we obtain a consistency condition for the asymptotic values of the bulk fields.  Some final remarks are given in section six.  An alternative  expression for the gauge invariant Scwarzian derivative is given in Appendix A.

\section{Preliminaries}

The Schwarzian derivative, which we shall denote by $S_t$, of a $C^3$ function $f(t)$ is defined as
\be S_tf =\frac{ f^{(3)}}{\dot f}-\frac 32 \left(\frac{\ddot  f}{\dot f}\right)^2\;,\label{theSchwzn}\ee
with the dots denoting ordinary derivatives of $f$ with respect to $t$ and  $f^{(3)}=\frac{d^3 f}{dt^3}$.  It is invariant under M\"{o}bius  transformations
\be  f\mapsto\frac{a f+b}{cf+d}\;, \qquad ad-bc\ne0\ .\label{frlitr}
\ee  For simplicity, we shall take $f$ to be  real.  We also restrict $a,b,c,d$ to being real and independent of $t$. The latter is the restriction to global transformations, which will be dropped in sections four and five.   By taking derivatives of (\ref{frlitr}), for instance
\beqa
 \dot f&\mapsto&\frac{ad-bc}{(cf+d)^2}\,\dot f\;,\label{gcircdotf}\eeqa
it is straightforward to verify that (\ref{theSchwzn}) is invariant under M\"obius transformations.

(\ref{frlitr}) defines the action of $GL(2,\mathbb{R})$ in  the  fractional linear representation ${\cal F}_{FL}$.  That is, for any $  f\in {\cal F}_{FL}$ and
\be g=
\begin{pmatrix}
a & b\cr c & d
\end{pmatrix}\in GL(2,\mathbb{R})\,,\;\quad {\rm det}\, g\ne 0\;,\label{gmatrix}\ee
the action  of $g$ on $f$ is
\be  g\,\triangleright f=\frac{a f+b}{cf+d}\ .\label{gl2ract}\ee
It is easily verified that the group product is preserved under (\ref{gl2ract})
\be [g'g]\,\triangleright f=g'\triangleright(g\,\triangleright f)\;,\qquad g,g'\in GL(2,\mathbb{R})\;.\label{grpprdt}\ee
 (\ref{gl2ract}) does not define a faithful representation of $GL(2,\mathbb{R})$ since the transformation  is unaffected by a continuous rescaling  of $g\rightarrow\lambda g\,,\lambda\ne0$.  There are two connected components of $GL(2,\mathbb{R})$ depending on the sign of the determinant of $g$. det $g>0$ corresponds to the proper transformations $GL^{+}(2,\mathbb{R})$, while   det $g<0$ yields the set of improper transformations, $GL^{-}(2,\mathbb{R})$. Only $GL^{+}(2,\mathbb{R})$ is a subgroup of $GL(2,\mathbb{R})$. Upon setting det$\,g=ad-bc=1$, (\ref{gl2ract}) gives the fractional linear representation of $SL(2,\mathbb{R})$.\footnote{More precisely, it is the coset space   $PSL(2,\mathbb{R})=SL(2,\mathbb{R})/Z_2$, since $g\,\triangleright f=(-g)\,\triangleright f$.}  We note from (\ref{gcircdotf}) that one consequence of this  is that the sign of $\dot f$ is preserved under $SL(2,\mathbb{R})$ transformations.

 $SL(2,\mathbb{R})$ induces translations, dilations and special conformal transformations on  elements of ${\cal F}_{FL}$.  In the defining representation, they are generated respectively by
\be T^0=\begin{pmatrix}
&-1\cr&
\end{pmatrix}\ , \qquad   T^1=\begin{pmatrix}
-\frac12&\cr &\frac 12
\end{pmatrix}\ , \qquad T^2=\begin{pmatrix}&\cr1&\end{pmatrix}\;.\label{sldefbas}\ee
$T^i,\;i=0,1,2,$ form a basis for the defining representation of the $sl(2,\mathbb{R})$ Lie algebra.  They satisfy the commutation relations:
\be [T^0,T^1]=T^0\ ,\qquad  [T^2,T^1]=-T^2\ ,\qquad  [T^0,T^2]=2T^1\;.\label{sl2ralg}
\ee
The nonvanishing traces of the product of two generators are $ {\rm tr} \,T^0T^2=-1$ and ${\rm tr}\, T^1T^1=\frac 12$, which can serve as components of a metric tensor on the $sl(2,\mathbb{R})$ Lie algebra, $\gamma^{ij}=2\,{\rm tr} T^iT^j$,
\be [\gamma^{ij}]=\begin{pmatrix} &&-2\cr&1&\cr-2 &&\end{pmatrix}\ .\label{mtrc}\ee
It will be useful to also introduce a  set of  dual basis vectors $T_i,\;i=0,1,2$.  They are defined using the inverse metric tensor $\gamma_{ij}$, $T_i=\gamma_{ij}T^j$.  Thus
\be T_0=-\frac{T^2}2\ ,\qquad  T_1=T^1\ ,\qquad T_2=-\frac{T^0}2\;.\label{dualbasis}\ee
One then  has
\be{\rm tr}\, T^iT_j=\frac 12\delta^i_j\ .\label{traceT}\ee
 The quadratic Casimir operator for $SL(2,\mathbb{R})$ is ${\cal C}_2=T^iT_i$, which has the value  $\frac 3 4$ in the defining representation.

We next construct the ``composite field'' ${\bf f}$ from $f$ and $\dot f$.  It  transforms under the global $SL(2,\mathbb{R})$ group in the adjoint representation.  First note that the M\"{o}bius transformation of  the $sl(2,\mathbb{R})$-valued function $T_0+f T_1+ f^2 T_2$ is  \begin{eqnarray}\label{SL2F}
 T_0+f T_1+ f^2 T_2 \mapsto \frac{ad-bc}{(cf+d)^2}\,g(T_0+f T_1+ f^2 T_2) g^{-1}\;.
\end{eqnarray}
If we then define
\be {\bf f:}=\frac {T_0+f T_1+ f^2 T_2}{\dot f}=-\frac 1{2\dot f}\begin{pmatrix} f&-f^2\cr 1& -f \end{pmatrix}\;,\quad \dot f\ne 0\label{boldF}\;,\ee  using   (\ref{gcircdotf}), we see that it transforms under the adjoint action of $SL(2,\mathbb{R})$
\begin{eqnarray}\label{Adjoint_f}
    \textbf{f}\mapsto g \textbf{f} g^{-1}.
\end{eqnarray}
$T_0+f T_1+ f^2 T_2$, and consequently ${\bf f}$, are singular and nilpotent,  ${\bf f}^2=0$.\footnote{It is also possible to construct ${\bf f}$ from spinors that transform under the fundamental representation of $SL(2,\mathbb{R})$. The spinors involve a square root of $\dot f$, so we have to be careful with signs. For $\dot f>0$, the fundamental spinor is $s=\frac 1{\sqrt{\dot f} }\begin{pmatrix} f \cr 1 \end{pmatrix}$.  From (\ref{frlitr}) and (\ref{gcircdotf}), it follows that $s\mapsto gs$, provided that  we make the restrictions: $ad-bc=1$ and $cf+d>0$.  Similarly, one can define  $\bar s=\frac 1{\sqrt{\dot f} }\begin{pmatrix} -1,f \end{pmatrix},\;\dot f>0$, which transforms with the inverse of the fundamental representation  of $SL(2,\mathbb{R})$, $\bar s\mapsto \bar sg^{-1}$, for $ad-bc=1$ and $cf+d>0$. Then the composite field ${\bf f}$  can be constructed from $s$ and $\bar s$ as follows:
$${\bf f}=\frac 12 s\bar s\;.$$
Because $\bar{s} s =0$, we immediately recover the nilpotency of ${\bf f}$.}
 Of course, the derivatives of $ {\bf f} $ also transform as in (\ref{Adjoint_f}), since  here we are  restricting to global transformations. $SL(2,\mathbb{R})$ invariants can be constructed from the trace of polynomials of ${\bf f}$ and  its derivatives. However, many such invariants  are trivial due to the nilpotency of $\textbf{f}$.  For example:
\be {\rm tr}\,{\bf f}^2={\rm tr}\,{\bf f}\dot {\bf f}={\rm tr}\,\dot{\bf f}\ddot {\bf f}= {\rm tr}\, {\bf f}^{(3)}{\bf f}=0\ ,\qquad\quad{\rm tr}\,\dot {\bf f}^2=-{\rm tr}\,{\bf f}\ddot{\bf f} =\frac 12\ .\label{simpids}\ee
The first set of non-trivial bilinear invariants are equal to the Schwarzian derivative,
\be  {\rm tr}\,\ddot {\bf f}^2= {\rm tr}\, {\bf f}^{(4)}{\bf f}=  S_t f\label{trfsq}\;,\ee
while others are derivatives of the Schwarzian derivative, such as:
\be {\rm tr}\, {\bf f}^{(4)}\dot{\bf f}=-\frac32 \,\frac d{dt} S_t f\ , \qquad \quad{\rm tr}\, {\bf f}^{(5)}{\bf f}=\frac 52 \,\frac d{dt} S_t f \ .\label{trfsq2} \ee
There exist many more identities involving  ${\bf f}$ and its derivatives.  One, which has some use in the next section, is
\be \ddot{\bf f}- [{\bf f},{\bf f}^{(3)}]+[\dot {\bf f}, \ddot{\bf f}] + 4 {\bf f}\,S_t f =0 \ .\label{cmplid}\ee
Lastly, we remark that $ \ddot {\bf f}$ is nilpotent when it is evaluated for $f=\frac{a t+b}{ct +d}$.  From (\ref{trfsq}) this leads to the well known result that the Schwarzian derivative is zero for such a function.

\section{ Schwarzian dynamics}
\setcounter{equation}{0}
\subsection{The ``free'' system}

$SL(2,\mathbb{R}) $ invariant dynamical systems have been constructed from  the Schwarzian derivative.  The associated Lagrangians naturally involve higher orders of derivatives (although they may be reformulated in terms of first order Lagrangians,\cite{Bagrets:2016cdf},\cite{Mertens:2017mtv} which we review at the end of this subsection).  In the simplest example, which we consider here, one can just set the Lagrangian equal to the Schwarzian derivative.  The resulting dynamics has been studied previously, see e.g., \cite{Galajinsky:2018ona}.  It was noted that such dynamics is different from the conformally invariant  system of de Alfaro, Fubini and Furlan.\cite{deAlfaro:1976vlx}

The  action here is
\be {\cal I}_0= -
\int dt \,S_tf= -
\int dt \, {\rm tr}\,\ddot {\bf f}^2\;,\label{Schactn}\ee where we used (\ref{trfsq}).   We shall call this the free system despite the nontrivial  self-interactions of $f$. Here $f$ is to be regarded as a fundamental degree of freedom,  and so variations of the action $ \delta{\cal I}_0$  should be carried with respect to this variable.   ${\bf f}$, on the other hand, is not a fundamental degree of freedom (hence the name of the composite field).  Arbitrary variations $\delta f$ in $f$ lead to the following variations in ${\bf f}$
\be \delta {\bf f}=\frac 1{\dot f}\left(\left(T_1+2f T_2\right)\delta f-{\bf f} \,\delta\dot f\right)=\frac 1{\dot f^2}\,\frac d{dt}\left(\dot f {\bf f}\right)\delta f\,-\,\frac 1{\dot f}\,{\bf f} \,\delta\dot f\;.\ee
The resulting variation of the action   (\ref{Schactn}) is
\beqa -\delta{\cal I}_0&=& 2
\int dt \, {\rm tr}\,\ddot {\bf f}\;\frac{d^2}{dt^2} \delta {\bf f}\cr&&\cr
&=&2\,\int dt \,\left(\frac d{dt}\left(\frac 1{\dot f} {\rm tr}\,{\bf f}\,{\bf f}^{(4)}\right)+\frac 1{\dot f^2}{\rm tr}\,{\bf f}^{(4)}\frac d{dt}\left(\dot f {\bf f}\right)\right)\,\delta f\;+\;B.T.\cr&&\cr
&=&2\,\int dt \, \frac 1{\dot f}\;{\rm tr}\left(2\,{\bf f}^{(4)}\dot{\bf f}+{\bf f}^{(5)}{\bf f}\right)\,\delta f\;+\;B.T.\;,\label{varyI0}\eeqa
with $B.T.$  denoting the boundary terms:
\be B.T.\,=\, 2\, {\rm tr}\left( \ddot{\bf f} \frac d{dt}\delta {\bf f}  -{\bf f}^{(3)}\delta {\bf f}  -{\bf f}\,{\bf f}^{(4)}\frac 1{\dot f}\delta f\right)\ .\label{BTforvarA}\ee
(\ref{varyI0}) leads to the following the equation of motion
\be  \frac 1{\dot f}\;{\rm tr}\left(2\,{\bf f}^{(4)}\dot{\bf f}+{\bf f}^{(5)}{\bf f}\right)
\,=\,0 \ .
\ee
Using the identities  (\ref{trfsq2}) this becomes
\be \frac 1{\dot f}\,\frac d{dt} S_t f =0\;,\label{eom}\ee
which for nonsingular $\dot f$ says that the Schwarzian derivative has a constant value for any solution.
The equation of motion can  equivalently be written as
\be\frac d {dt}\left(\frac{ f^{(3)}}{\dot f^2}- \frac{\ddot  f^2}{\dot f^3}\right)=0\;.\label{Seqomot}\ee

Solutions to the equation of motion can be found from an old property of the Schwarzian derivative, which says that the potential  function $q(t)^2$ in the Sturm-Liouville equation
\be \ddot \phi(t)-\frac 14q(t)^2\phi(t)=0\label{secordeq}\ee
can be reconstructed from the solutions for $\phi$.\cite{Ovsienko,Osgood1998}  Namely, $q(t)^2=-2 S_t\left(\frac {\phi_2}{\phi_1}\right)$, where $\phi_1$ and $\phi_2$ are linearly independent solutions of  (\ref{secordeq}). As our equation of motion (\ref{eom}) fixes  the Schwarzian derivative to be constant, the relevant function $q(t)^2$ is a constant, which we simply denote by $q^2$, reducing the Sturm-Liouville equation (\ref{secordeq}) to the harmonic oscillator equation (with imaginary frequency, if $q^2>0$).  The ratio of linearly independent solutions, $f=\frac{\phi_1}{\phi_2}$, will solve (\ref{eom}). A general solution to (\ref{secordeq}) is given by $\phi = a e^{\frac{q}{2}t} + b e^{-\frac{q}{2}t}$ and the two solutions, $\phi_1 = a e^{\frac{q}{2}t} + b e^{-\frac{q}{2}t}$ and $\phi_2 = c e^{\frac{q}{2}t} + d e^{-\frac{q}{2}t}$, will be linearly independent if and only if $ad-bc \ne 0$.\footnote{For the case of $q^2 <0$, one has to take the real linear combinations to guarantee that $f$ is real.} So, the general solution to $q^2=-2 S_t\left(f\right)$ is
\be f=\frac {ae^{\frac{q}{2}t}+be^{-\frac{q}{2}t}}{ce^{\frac{q}{2}t} +de^{-\frac{q}{2}t}}\label{mbssoleqt}\;.\ee

One can also obtain a general solution to (\ref{Seqomot}) by performing a M\"{o}bius transformation of an obvious particular solution,  the exponential function $e^{rt}$, $r=const$:
\be f=\frac {ae^{rt}+b}{ce^{rt} +d}\label{mbssoleqt1}\;,\ee
where $a,b,c,d$ are constants.
 It  can be easily checked that the on-shell value of the Schwarzian for solution  (\ref{mbssoleqt1}) is $-\frac{r^2}2$, so (\ref{mbssoleqt}) and (\ref{mbssoleqt1}) are the same for $r=q$ and the condition that $ad-bc \ne 0$ is equivalent to the requirement of the non-degeneracy of the corresponding M\"{o}bius transformation. We shall  argue that  (\ref{mbssoleqt}) is, in fact, the most general solution to (\ref{eom}) at the end of this subsection.\footnote{Because the Schwarzian derivative of any M\"{o}bius transformation is identically zero, there should be another family of solutions, namely $f=\frac{At + B}{Ct + D}$, which looks different from (\ref{mbssoleqt1}). This family of solutions corresponds to another particular solution to (\ref{eom}) or (\ref{Seqomot}).  An example is $f = t$ (or $f = \frac{1}{t}$, which is related to $t$ by an improper M\"{o}bius transformation). But, in fact, this family is a limiting case of the family (\ref{mbssoleqt1}). By setting in (\ref{mbssoleqt1}) $a=\frac{A}{r}$, $b=B - \frac{A}{r}$, $c=\frac{C}{r}$ and $d= D - \frac{C}{r}$ and taking the limit $r \rightarrow 0$, we get what is claimed. So, there is no need to consider these two families separately. Moreover, we note that $f=\frac{At + B}{Ct + D}$ are a three-parameter set of zero mode solutions of the action (\ref{Schactn}), which, as stated above, easily follows from the nilpotency of $\ddot {\bf f}$ when evaluated for such solutions.}

Noether charges associated with the $SL(2,\mathbb{R})$ symmetry of the action (\ref{Schactn}) can be computed from the boundary terms (\ref{BTforvarA}). The infinitesimal version of the $SL(2,\mathbb{R})$ symmetry transformations in (\ref{frlitr}) is
\be f\rightarrow f -\epsilon_0-\epsilon_1f-\epsilon_2f^2\;,\label{symff}\ee
where $\epsilon_k, k=0,1,2,$ are infinitesimal parameters associated with translations, dilations and special transformations, respectively.
Since here $\epsilon_k$ are independent of $t$, one gets  simple transformations for $\dot f$,  $\dot f\mapsto \dot f -(\epsilon_1 +2\epsilon_2  f)\dot f$.  Then the resulting variation of the adjoint vector ${\bf f}$  is
\be  \delta {\bf f}= [\epsilon_k T^k,{\bf f} ]\ .\label{symfbldf}\ee
Substituting (\ref{symff}) and (\ref{symfbldf}) into (\ref{BTforvarA}), using $B.T.=-N^i\epsilon_i$, gives the following  Noether charges
\beqa N^i&=&-2 \,{\rm tr}\left(   T^i\left([\dot {\bf f}, \ddot{\bf f}] - [{\bf f},{\bf f}^{(3)}] \right)+\frac {f^i}{\dot f}{\bf f}\,{\bf f}^{(4)}\right)\cr&&\cr
&=&2 \,{\rm tr}  T^i\left( \ddot{\bf f}+ 4 {\bf f}\,S_t f\right)-2\,\frac {f^i}{\dot f} \,S_tf\;,\label{Nother}\eeqa
where we used the identities (\ref{trfsq}) and  (\ref{cmplid}).
Upon evaluating them explicitly one gets\footnote{These charges are, for the most part, the same as the constants of motion found in \cite{Galajinsky:2018ona}.}
\beqa N^0&=&\frac{ f^{(3)}}{\dot f^2}- \frac{\ddot  f^2}{\dot f^3}\ ,\cr&&\cr
N^1&=&N^0 f -\frac{\ddot f}{\dot f}\ ,\cr&&\cr
N^2&=& N^0 f^2-2\,\frac{\ddot ff}{\dot f}+2\dot f\ . \label{Notcha} \;\eeqa
They are associated, respectively, with the  translation, dilation and special conformal symmetries. The conservation of $N^0$ obviously follows directly from the equation of motion (\ref{Seqomot}). 
It can be checked that when evaluated on-shell,  the Noether charges depend only on the constants parametrizing the solutions.  For the generic solution (\ref{mbssoleqt}), their values are $N^0=-\frac{2cdq}{ad-bc}$, $N^1=-\frac{(ad+bc)q}{ad-bc}$, $N^2=-\frac{2abq}{ad-bc}$.

One can form a vector in the  $sl(2,\mathbb{R})$ Lie algebra out of the three Noether charges (\ref{Nother}), namely  ${\bf N}=N^i T_i$.  Using the identity ${\bf L}=2T_i\, {\rm tr}\,T^i{\bf L}$, for any ${\bf L}\in sl(2,\mathbb{R})$, along with the definition (\ref{boldF}), it simplifies to
\beqa {\bf N}=N^i T_i&=&\ddot{\bf f}+2\, {\bf f}\,S_t f \ .\label{bfN}
\eeqa
 Since  ${\bf N}$ is a linear combination of ${\bf f}$ and $\ddot{\bf f}$ with invariant coefficients, it undergoes a similarity transformation under the action of  $SL(2,\mathbb{R})$
\be {\bf  N}\mapsto g {\bf  N} g^{-1} \;,\qquad g\in SL(2,\mathbb{R})\ .\label{simtrfrN} \ee
Upon expanding this in terms of the three Noether charges (\ref{Notcha}) one gets
\be g\,\triangleright \begin{pmatrix}N^0\cr   N^1\cr  N^2\end{pmatrix}=\frac1{ad-bc}\begin{pmatrix} d^2 &2cd & c^2\cr bd &ad+bc &ac\cr b^2 &2 a b&a^2\end{pmatrix} \begin{pmatrix} N^0\cr N^1\cr N^2\end{pmatrix}\;,\label{Ntrns}\ee
where $g$ is  defined as in (\ref{gmatrix}).
Here we have not restricted the determinant of $g$ to be one, but we see once again that the transformations are  not affected by an overall rescaling of $g$, and so  the determinant of $g$ acts trivially, as  should be the case.

We finally relate the above discussion to the first order  Lagrangian formulation of Schwarzian dynamics introduced in  \cite{Bagrets:2016cdf}, and further discussed in \cite{Mertens:2017mtv}.   This formulation can be obtained by first
integrating the Schwarzian action (\ref{Schactn}) by parts (and now throwing away the boundary terms) to give the positive integral:
\be  \frac 12\int dt \left(\frac{\ddot  f}{\dot f}\right)^2\ .\ee  The solutions (\ref{mbssoleqt}) or (\ref{mbssoleqt1})   found previously to the equations of motion for  $f$ were M\"{o}bius transformations  of the exponential function. After restricting $ad-bc=1$, they become  $SL(2,\mathbb{R})$ transformations of the exponential function.
We recall from (\ref{gcircdotf}) that  the sign of $\dot f$ is preserved under  $SL(2,\mathbb{R})$ transformations.
Let us  choose  $\dot f>0$ by identifying it with an exponential function $e^{x(t)}$.  This can be implemented in the action integral with the  use of a Lagrange multiplier $\lambda=\lambda(t)$.  We then end up with the following first order Lagrangian
\be {\cal I}_0'=\int dt \left( \frac{\dot  x^2}{2}+\lambda(e^x-\dot  f)\right)\ .\label{calIzer}\ee
Here $x$ and $f$ are to be regarded as the fundamental degrees of freedom of the system.
The  equation of motion for $x$ reveals an exponential force
\be \ddot x=\lambda e^x\;,\label{eomfrx}\;\ee  while the equation of motion for $f$ says that $ \lambda$ is a  constant.  In fact, from (\ref{Notcha}) and (\ref{eomfrx}), $\lambda$ is the Noether charge $N^0$.  (The other two  Noether charges $N^1$ and $N^2$ are nonlocal functions of $x(t)$.)  We note that an exponential force term also occurs in Liouville theory with a boundary,\cite{Mertens:2018fds} and in the context of black holes.\cite{Lin:2019kpf}
Given some initial values for $x$ and its $t$-derivative,  $x(0)$ and $ v(0)= \dot x|_{t=0}$,  the general solution to (\ref{eomfrx}) is
\be x(t)=x(0) +qt -2\log\left(1+e^{qt}+\frac{v(0) }q\left(1-e^{qt}\right)\right)+\log 4\;,\label{gnrlsol}\ee
where here $q$ is defined by $q^2=v(0)^2-2\lambda e^{x(0)}$.   This solution  agrees with our previous solution  (\ref{mbssoleqt}).  To see this, use  the identification of $\dot f$ with $e^x$,
and then integrate the exponential of (\ref{gnrlsol}) to obtain the solution for $f(t)$
\be\int^t e^{x(t')} dt'=\frac{(q-v(0))C\,e^{qt}\;+\;(q+v(0))C -\frac{4qe^{x(0)}}{q-v(0)}}{(q-v(0))\,e^{qt} \;+\;q+v(0)}\;,\label{intex}
\ee
where $C$ is an integration constant.  This has the same form as  (\ref{mbssoleqt}), and furthermore one can relate the three independent constants amongst $a,b,c$ and $d$ in  (\ref{mbssoleqt}) to the integration constants $x(0)$, $v(0)$ and $C$ in (\ref{intex}). Since (\ref{gnrlsol}) is the most general solution to the equation of motion (\ref{eomfrx}), it follows that  (\ref{mbssoleqt}) is the most general solution to the equation of motion (\ref{Seqomot}).

\subsection {Interactions}

Next let us consider going beyond the ``free'' theory.  Here it is most convenient to describe the ``free'' system by the  Lagrangian given in (\ref{Schactn}), which is written in terms of higher order derivatives,  rather than that of (\ref{calIzer}).  This is because the latter formulation generally leads to nonlocal interactions.
A natural way to include  $SL(2,\mathbb{R})$ invariant interactions  is to couple the Noether charges in ${\bf N}$ to  some $sl(2,\mathbb{R})-$valued  fields ${\bf A}=A_i T^i$.  In order to ensure invariance under (global) $SL(2,\mathbb{R})$  transformations we should demand that ${\bf A}$, like ${\bf N}$, transforms in the adjoint representation
\be {\bf A}\mapsto g{\bf A} g^{-1}\ , \label{adactonA} \ee
so that the trace of products with ${\bf A}$ is invariant.
In terms of the components $A_i$ one has
\be g\,\triangleright \begin{pmatrix}A_0\cr   A_1\cr  A_2\end{pmatrix}=\frac1{ad-bc}\begin{pmatrix} a^2 &-ab & b^2\cr -2ac& ad+bc&- 2bd\cr c^2&-cd& d^2\end{pmatrix}\begin{pmatrix}A_0\cr A_1\cr A_2\end{pmatrix}\ .\label{trnsa012}\ee
Note that this differs from the analogous transformation of the Noether charges (\ref{Ntrns}).  This is due to the fact that $A_i$ and $N^i$ are components of dual basis vectors.
We now easily construct an $SL(2,\mathbb{R})$ invariant coupling term
\be2\,{\rm tr}\,{\bf  N}{\bf A}=N^i A_i\;,\label{couptrm}\ee
where we utilized trace identity (\ref{traceT}).
Additional  $SL(2,\mathbb{R})$ invariant terms  should be introduced in the action in order to obtain  nontrivial dynamics, since otherwise the equations of motion resulting from variations of $A_i$ in the total action would lead to $N^i=0$. An obvious example would be a kinetic energy term for ${\bf A}$: $\;$tr$\,\dot {\bf A}^2=\frac 12\dot A_i \dot A_j\gamma^{ij}=\frac 12(\dot  A_1)^2-2\dot A_0\dot A_2\;,$  where the metric tensor was defined in  (\ref{mtrc}).
Many more $SL(2,\mathbb{R})$-invariant terms can be constructed from ${\bf A}$ and ${\bf N}$.  For example, one can introduce  the quadratic terms: tr${\bf A}^2{\bf N}^2$, tr$({\bf A}{\bf N})^2$,  tr${\bf A}^2\,$tr${\bf N}^2$, $($tr${\bf A}{\bf N})^2$.  We can  also include  analogous terms  involving derivatives of ${\bf A}$.  Moreover, we can  couple ${\bf A}$ directly to ${\bf f}$ defined in (\ref{boldF}), as ${\bf f}$ also transforms in the adjoint representation of  $SL(2,\mathbb{R})$.

The gauge invariant generalization of the Schwarzian derivative, which we examine in the next section, contains  $sl(2,\mathbb{R})$ gauge potentials, which we shall also denote by ${\bf A}$.  Upon restricting to global transformations ${\bf A}$ transforms according to (\ref{adactonA}), or (\ref{trnsa012}).
We shall show that (\ref{couptrm}) is recovered from the  gauge invariant version of the Schwarzian derivative at first order in the expansion of ${\bf A}$.  At second order we obtain a linear combination of the terms:
 ${( {\rm tr}\, {\bf f}\dot  {\bf A})^2},\,{\rm tr}({\bf f} {\bf A}\dot{\bf A})$, $ ({\rm tr}\, {\bf f}{\bf A})^2$ and tr${\bf A}^2$.

\section{$SL(2,\mathbb{R})$ gauge symmetry}
\setcounter{equation}{0}
\subsection{The gauge invariant Schwarzian derivative}

Here we promote the global $SL(2,\mathbb{R})$ transformations   to local (or gauge) transformations.
So now $g$ (with det$\,g=1$) in (\ref{gmatrix}) is a function of $t$. The Schwarzian derivative (\ref{theSchwzn}) is not  invariant under such transformations.  This is obvious  since (\ref{gcircdotf}) no longer holds.
 Instead, we have
\begin{eqnarray}\label{SL2fdot_local}
    \dot{f}\mapsto  \frac{1}{(cf+d)^2}\dot{f}\left(1 -2 \mathrm{tr} \left( \dot{g} \textbf{f} g^{-1} \right)\right) \;.
\end{eqnarray}
 In order to obtain invariance under local transformations it will be necessary to replace $t-$derivatives with some sort of covariant analogs.
This requires the construction of two objects: 1) the counterpart to the composite field $\textbf{f}$, which we denote by ${\bf f}_A$, that  transforms in the adjoint representation of $SL(2,\mathbb{R})$, {\it  even under the local  transformations}, and 2) the covariant analog of the time derivative of $\textbf{f}_A$.  The first step introduces the covariant derivative $D_A$ acting on
the fractional linear representation, while the second requires the covariant derivative ${\bf D}_A$ acting on
the adjoint representation.

Concerning 1), it is clear that we  need to deform $\textbf{f}$. This is because while $T_0+f T_1+ f^2 T_2$ still transforms as in (\ref{SL2F}) (due to the absence of derivatives), (\ref{Adjoint_f}) does not hold  for  local $SL(2,\mathbb{R})$ transformations (due to (\ref{SL2fdot_local})).
So, we have to compensate for the extra term in (\ref{SL2fdot_local}).
Towards this end, and for the purpose of 2), we introduce an $sl(2,\mathbb{R})$-valued gauge field $ {\bf  A} = A_k T^k $, which is an extension of the vector field of the previous section.  It should transform under local $SL(2,\mathbb{R})$ according to
\begin{eqnarray}\label{Gauge}
 {\bf  A}\mapsto g{\bf A}g^{-1} + \dot{g}g^{-1}\;,
\end{eqnarray}
which is the extension of  (\ref{adactonA}).
  With the help of this field, we can define the covariant time derivative $ D_A$ acting on $f$,
\begin{eqnarray}\label{CovDer1}
    D_A f := \dot{f} +A_0+ A_1f + A_2 f^2 \equiv \dot{f}(1 + 2 \mathrm{tr} (\textbf{Af})) \;.
\end{eqnarray}
Using (\ref{SL2F}), (\ref{SL2fdot_local}) and (\ref{Gauge}), it is a one-line exercise to verify that $D_A f$ transforms covariantly, i.e., as in (\ref{gcircdotf})
\begin{eqnarray}
    D_A {f}\mapsto \frac{1}{(cf+d)^2}D_A{f}.
\end{eqnarray}
Then we can easily define the \textbf{gauged} covariant version of the composite field
  \begin{eqnarray}\label{Composite_covariant}
    \textbf{f}_A := \frac{T_0+f T_1+ f^2 T_2}{D_A f}  =\frac{1}{1 + 2 \mathrm{tr} (\textbf{Af})} \textbf{f}\;,
\end{eqnarray}
which transforms  in the adjoint representation under the gauge transformations:
\begin{eqnarray}
    \textbf{f}_A\mapsto g \textbf{f}_A g^{-1} .
\end{eqnarray}
$  \textbf{f}_A$, like ${\bf f}$, is singular and nilpotent,  ${\bf f}_A^2=0$.
Note that the  ``form-factor'' $\frac{1}{1 + 2 \mathrm{tr} (\textbf{Af})}$ in the definition (\ref{Composite_covariant}) allows  for easy control over the expansion in ${\bf A}$ about the undeformed composite field ${\bf f}$,
\begin{eqnarray}\label{Composite_covariant_expansion}
    \textbf{f}_A = \textbf{f}\sum\limits_{n=0}^{\infty}\left(-2 \,\mathrm{tr} \left(\textbf{Af}\right)\right)^{n}.
\end{eqnarray}

Concerning 2), the procedure  is now trivial. The covariant derivative in the adjoint representation is implemented using the commutator with ${\bf A}$.  So acting on $\textbf{f}_A$ we get:
\begin{eqnarray}\label{CovDer2}
    {\bf D}_A \textbf{f}_A := \dot{\textbf{f}}_A - [{\bf A},{\bf{f}}_A].
\end{eqnarray}
(For simplicity, the same gauge field ${\bf A}$ is assumed in both covariant derivatives, (\ref{CovDer1}) and (\ref{CovDer2}).)
Then it is a standard, easily verifiable, fact that ${\bf D}_A {\bf f}_A$ gauge transforms in the adjoint representation, i.e., covariantly:
\begin{eqnarray}\label{Covariant1}
{ \bf D}_A \textbf{f}_A\mapsto   {\bf D}_{A'} \textbf{f}'_{A'} = g ({\bf D}_A \textbf{f}_A ) g^{-1}. \end{eqnarray}
It can be checked that all of the identities (\ref{simpids}) also hold for ${\bf f}_A$ upon  replacing the ordinary derivatives by covariant ones ${\bf D}_{A}$.

Now we easily write the gauge invariant version of the Schwarzian derivative (\ref{theSchwzn})\beqa   S[A]_t (f):&=&  \mathrm{tr} \left(\mathbf{D}_A\mathbf{D}_A\textbf{f}_A\right)^2\cr&&\cr&=&{\rm tr }\left( \ddot{\textbf{f}}_A - [\dot{\bf A},{\bf{f}}_A]-2[{\bf A},\dot{\bf{f}}_A ]+[{\bf A},[{\bf A},{\bf f}_A]]\right)^2\label{Schw_action_local}.
\eeqa
In Appendix A we give an equivalent expression for (\ref{Schw_action_local}) that is written explicitly in terms of $f$ and $A_k$.
Of course, (\ref{Schw_action_local}) reduces to the Schwarzian derivative (\ref{theSchwzn}) when ${\bf A}=0$, $ {\cal S}{[0]}_tf ={ S}_tf$.  Using  (\ref{Composite_covariant_expansion}), we can expand (\ref{Schw_action_local})  about the  Schwarzian derivative to all orders in  ${\bf A}$.  After defining $\xi:=-2 \,\mathrm{tr} \left(\textbf{Af}\right)$, the expansion up to second order is
\beqa   S[A]_t (f)&=&{\rm tr}\Biggl(\ddot{\bf f}\left(1+\xi+\xi^2\right)+2\dot {\bf f}\dot \xi(1+2\xi)+{\bf f}(\ddot \xi +2 \dot \xi^2+2\xi\ddot \xi)\cr&&\cr&&- [\dot{\bf A},{\bf{f}}](1+\xi)-2[{\bf A},\dot{\bf{f}} ](1+\xi)-2[{\bf A},{\bf f}]\dot \xi+[{\bf A},[{\bf A},{\bf f}]]\Biggr)^2+{\cal O}({\bf A}^3)\cr&&\cr&=& S_t (f)+ S^{(1)}_t (f)+ S^{(2)}_t (f)+{\cal O}({\bf A}^3)\;,
\eeqa
where
\beqa   S^{(1)}_t (f)&=&2\,{\rm tr}\ddot {\bf f} \left(-2 [{\bf A},\dot{\bf{f}} ]- [\dot{\bf A},{\bf{f}}]+\frac{ d^2}{dt^2}(\xi{\bf f}
)\right)\ ,\cr&&
\cr S^{(2)}_t (f)&=&2\,{\rm tr}\ddot {\bf f}\biggl([{\bf A},[{\bf A},{\bf f}]]- 2\frac d{dt}\left( \xi[{\bf A},{\bf{f}}]\right)+ \xi[\dot{\bf A},{\bf{f}}]+2\xi\frac{ d^2}{dt^2}(\xi{\bf f}
)
-\xi^2\ddot {\bf f}+2{\bf f} \dot \xi^2\biggr) \cr&&\cr&&+{\rm tr}\left(-2[{\bf A},\dot{\bf{f}} ] - [\dot{\bf A},{\bf{f}}]+\frac{ d^2}{dt^2}(\xi{\bf f}
)\right)^2 \ .\eeqa
Up to total derivative terms, the first order correction is just the linear coupling to the Noether charges (\ref{couptrm})
\be S^{(1)}_t (f)=2\,{\rm tr}\,{\bf  N}{\bf A}\,+\, {\rm total}\;{\rm derivatives.}\label{lovarnA}\ee
This can be verified by using (\ref{trfsq}), (\ref{cmplid}) and (\ref{bfN}). After some work the second order correction can be shown to simplify to
\be S^{(2)}_t (f)=-2{( {\rm tr}\, {\bf f}\dot {\bf  A})^2}+4\,{\rm tr}({\bf f A}\dot {\bf A})-4\, ({\rm tr}\, {\bf f A})^2 \,{\rm tr}\,\ddot {\bf f}^2\,-\,2\,  {\rm tr}{\bf  A}^2\,+\, {\rm total}\;{\rm derivatives.}\quad\label{scndordnA}\ee

The Schwarzian action (\ref{Schactn}) can be promoted to its gauge invariant counterpart
\be {\cal I}[A]= -
\int dt \,S_t[A](f)= -
\int dt \, {\rm tr}\,\left(\mathbf{D}_A\mathbf{D}_A\textbf{f}_A\right)^2\;,\label{ginvactn}\ee
with both $f$ and ${\bf A}$ denoting  degrees of freedom.  However, the latter   degrees of freedom can be gauged away if the domain is topologically  trivial, i.e., $\mathbb{R} $ (with unrestricted boundary values at $t\rightarrow \pm\infty$).  In this case a local $SL(2,\mathbb{R})$ transformation by
\be g^{(0)}(t) = g(-\infty){\rm T} \,e^{-\int_{-\infty}^t {\bf A}(t')dt'}\;,\ee
will map ${\bf A}(t)$ to zero.   Here T denotes time ordering with earlier times to the left. In this case (\ref{ginvactn}) is equivalent to the Schwarzian action  (\ref{Schactn}), or the latter is a gauge choice of  (\ref{ginvactn}).

\subsection{Nontrivial topological sectors}

 The situation is much more interesting when the domain is $S^1$ (or equivalently, $ \mathbb{R}$ with periodic boundary condition).  For this let us first do a Wick rotation $t\rightarrow i\tau$ and take  $0\le \tau<2\pi$. Now the trivializing gauge transformations belong to  distinct homotopy classes $\{  G^{(n)},\;n=0,\pm1,\pm 2,...\}$.  This comes from the requirement that $g^{(n)}(2\pi)=g^{(n)}(0)$,  $g^{(n)}\in G^{(n)}$, and hence the  holonomy
$ {\rm T}\,e^{-i\oint {\bf A}(\tau)d\tau} $ is the identity. The homotopy classes are associated with the fundamental group $\pi_1\left(SL(2,\mathbb{R})\right)=\mathbb{Z}$. Say that $  G^{(0)}$ contains the identity $g=\BI$.  So $  G^{(0)}$ is the set of gauge transformations that are connected to the identity. To write down the winding  number that labels  the homotopy class, we can first perform a gauge transformation on ${\bf A}$ by an $h^{(0)}\in G^{(0)}$ which takes it everywhere to the $T^1$ direction (which generates the $U(1)$ subgroup $SL(2,\mathbb{R})$),
\be{\cal A}\, T^1=h^{(0)}{\bf A}{h^{(0)}}^{-1} -i \frac d{dt}\left({h^{(0)}}\right){h^{(0)}}^{-1}\ .\ee
Then the winding number is
\be n=\frac 1{2\pi}\oint d\tau\, {\cal A}=\frac 1\pi\oint d\tau\,{\rm tr} \,T_1\left(h^{(0)}{\bf A}{h^{(0)}}^{-1}-i \frac d{dt}\left({h^{(0)}}\right){h^{(0)}}^{-1}\right)\ .\label{windno}\ee
Of course $h^{(0)}$ is not unique.  Under `small' gauge transformations, i.e.,  $ g^{(0)}\in G^{(0)}$, it transforms as $h^{(0)}\mapsto h^{(0)}{g^{(0)}}^{-1} $, which along with ${\bf A} \mapsto g^{(0)}{\bf A}{g^{(0)}}^{-1} -i \frac d{dt}\left( g^{(0)}\right){g^{(0)}}^{-1}$, leaves $n$ invariant.  (We note that the expansion in ${\bf A}$ carried out above cannot be performed when $n\ne 0$.)
A simple gauge choice  for winding number $n$
is ${\bf A}_{(n)}=-2 n T^1$.  The evaluation of  (\ref{Schw_action_local}) in this gauge gives
\be  S[{\bf A}_{(n)}]_t (f)=\frac{f^{(3)}+4  n^3f}{\dot f-2 n f}+\frac{-\frac 32\ddot f^2+6n\dot f \left(
  \ddot  f -2n \dot f+2  n^2f \right)}{ \left(\dot f-2 n f\right)^2}\ . \ee
  This result is invariant under global $SL(2,\mathbb{R})$ only for $n=0$, confirming the uniqueness of the Schwarzian derivative.  For $n\ne 0$ the global symmetry is broken to $U(1)$, as easily follows from (\ref{adactonA}).
 The conclusion in this case is that there are infinitely many vacua labeled by $n\in \mathbb{Z}$, but only one, which is invariant under global $SL(2,\mathbb{R})$. The vacua are connected by ``large'' gauge transformations, i.e., $g^{(m)}\in G^{(m)},$ where $m$ is a nonzero integer.  For example, if we  take $g^{(m)}=e^{-2im\tau\, T^1}$, which is globally well defined on $S^1$, then it gauge transforms        ${\bf A}_{(n)}=-2 n T^1$ to  ${\bf A}_{(n+m)}=-2( n +m)T^1$.   The large gauge transformations resemble the instantons of QCD.  The unitary representation of any  $g^{(1)}\in G^{(1)} $ in the quantum theory is a phase $e^{i\theta}$, for some real $\theta$, analogous to the $\theta$ angle of QCD. 
 
 We note that while the gauge transformations are classified by an integer, $n$ as defined  on the space of $SL(2,\mathbb{R})$ potentials in (\ref{windno}) need not be an integer. Let's now call it $\alpha$. For non-integer values of $\alpha$, we cannot gauge transform ${\bf A}$ to zero using a large gauge transformation or one that is connected to the identity.  Therefore,  ${\bf A}$ with $\alpha\notin \mathbb{Z}$ does not describe a vacuum configuration. A gauge choice for  non-vacuum configurations can  be given by  ${\bf A}_{(\alpha)}=-2 \alpha T^1$, here with $\alpha\notin \mathbb{Z}$.  On the other hand, by applying a large gauge transformation we can change the value of $\alpha$  in (\ref{windno}) by an integer $m$.  Thus the  non-vacuum configurations form  equivalence classes  with non-integer values of $\alpha$ mod $\mathbb{Z}$.  They then  can be  labeled by  $0<\alpha<1 $. The representative of each equivalence class can be chosen as ${\bf A}_{(\alpha ,m)}=-2 (\alpha + m) T^1$ and for $0<\alpha<1 $ can be thought as gauge field ``excitations'' above the vacuum in $m$-th sector, considered above.

\section{Application to JT gravity}
\setcounter{equation}{0}

JT gravity is a theory of gravity in two space-time dimensions.\cite{T1983,J1985}
There are two main approaches to the subject: the original metric tensor  formulation  and the gauge theory formulation.\cite{FUKUYAMA1985259,Isler,CHAMSEDDINE198975,Jackiw:1992bw,Celada:2016jdt,Pinzul:2024zkl}  Here we shall only be concerned with the  latter.
In the gauge theory approach  the dynamical degrees of freedom on the two dimensional {\it bulk} $D$ consist of a space-time scalar and a connection one form, both of which are valued in the $sl(2,\mathbb{R})$ Lie algebra. We denote the former by ${\bf B}$. It transforms in the adjoint representation under $SL(2,\mathbb{R})$ gauge transformations
\beqa {\bf B}\mapsto {\bf  B}'&=& g {\bf B} g^{-1}\ .\eeqa
For the latter, we  simply extend the connection  ${\bf A}$ of the previous section from the one-dimensional domain, which we here regard as the boundary $\partial D$ of $D$, to $D$.
 The extension of gauge transformations
 (\ref{Gauge}) to $D$ is
\beqa {\bf A}\mapsto {\bf  A}'&=& g {\bf A}g^{-1} + dg  g^{-1}\label{sl2rgts}\;,\eeqa
where now $g$ is an $SL(2,\mathbb{R})$-valued function on $D$.
One can introduce  the $SL(2,\mathbb{R})$ curvature on $D$ (which of course doesn't exist on $\partial D$)
\be {\bf F}=d{\bf A}-{\bf A}^2\;,\label{sl2rcrv}\ee which transforms in the adjoint representation   ${\bf F}\mapsto {\bf F}'= g {\bf F} g^{-1}$  under gauge transformations.

The  dynamics of ${\bf A}$ and ${\bf B}$ on $D$ are determined by a BF action
\be
S_{JT}=\int_{D} \,{\rm tr}\,{\bf BF}\;.\label{actnltlab}
\ee (\ref{actnltlab}) is  gauge invariant, and the resulting equations of motion are
\be
{\bf F}=0\;,\qquad d{\bf B}+[{\bf A,B}]=0\;.\label{wontoo3}
\ee
Solutions  for the connections are  pure gauges
 \be {\bf A}= dh h^{-1}\;,\;\;\quad h\in SL(2,\mathbb{R})\label{pgponH2}\;,\ee
while solutions for  the space-time scalar  are ${\bf B}=h \ell h^{-1}$, $\ell$ being an arbitrary constant vector in the $sl(2,\mathbb{R})$ Lie algebra.

The Lie algebra components of ${\bf A}$ are associated with zweibein  one forms $e^a$, $a=0,1$ and the spin connection $\omega$, while the resulting  Lie algebra components  for ${\bf F}$ consist of the spin curvature $R$ and torsion two-forms $\theta^a, a=0,1$, see, e.g., \cite{Pinzul:2024zkl} for more details. The metric tensor is constructed, as usual, from the zweibein one forms,
\be
g_{\mu\nu}[V]= \eta_{ab}e^a_\mu e^b_\nu\;,\label{geiei}
\ee $\eta_{ab}$ being a flat metric and the indices $\mu,\nu=0,1$ specify space-time components.
 The choice for the decomposition into zweibeins and the spin connection is not unique, and it determines the resulting space-time. An example is \be {\bf A}=\frac 12 (\omega+e^1)T^0-e^2 T^1+\frac 12 (\omega-e^1)T^2\;,\ee
and the resulting $SL(2,\mathbb{R})$ curvature two form (\ref{sl2rcrv}) is
\be {\bf F}=\frac 12(R+e^1 e^2+\theta^1)T^0 -\theta^2 T^1+\frac 12(R+e^1 e^2-\theta^1)T^2\;,\ee
where the spin curvature two form is $R=d\omega $ and the  torsion is given by $ \theta^1=de^1+\omega e^2\,$ and  $\theta^2=de^2-\omega e^1$.  So here the zero curvature condition for ${\bf F}$  gives $R=-e^1e^2$ and $\theta^a=0$, implying   torsionless anti-de Sitter space.

It is well known that boundary terms and/or boundary conditions must be introduced in order that the  total action  is everywhere differentiable. This is true in the gauge theory formulation, as well as in the metric formulation.  In the gauge theory formulation, one obtains the boundary term \be\int_{\partial D} \,{\rm tr}\,{{\bf  B}\delta{\bf  A}}\label{btfrmBF}\;,\ee as a result of arbitrary variations of ${\bf A}$ in the action (\ref{actnltlab}), \footnote{ The presence of a boundary term would lead to a singular functional derivative localized on the boundary. Thus to have a regular functional derivative one should remove the boundary term.} which  can be removed either by imposing boundary conditions, or with the introduction of a boundary term in the action, or both. In the metric theory formulation, the boundary terms resulting from the variations of the metric tensor in  the JT gravity action, as well as in the Einstein Hilbert action, are standardly removed using the introduction of  the Gibbons-Hawking-York term.\cite{York:1972sj,Gibbons:1976ue} (It should be noted that the boundary action is not unique. For a discussion of the non-uniqueness of boundary  conditions for gravity see, for example, \cite{Dyer:2008hb}. For the specific case of JT gravity, see \cite{Goel:2020yxl}.) Boundary conditions can be further imposed in order to  specify the boundary dynamics. A particular choice of boundary condition was found for JT gravity, whereby a  Schwarzian action results  from the Gibbons-Hawking-York term.\cite{Maldacena:2016upp}
With a Schwarzian boundary action, the  local symmetry of the bulk is broken down to a global $SL(2,\mathbb{R})$ symmetry at the boundary, the latter symmetry being the central topic of sections 2 and 3 of this paper.
From section 4 we saw how this symmetry could be extended to a local  $SL(2,\mathbb{R})$ symmetry.  This  suggests the possibility of enlarging the global  symmetry at the boundary  to a gauge symmetry.  So let's consider the possibility of replacing the Schwarzian action at the boundary by the gauge invariant one (\ref{ginvactn}).  The boundary variation (\ref{btfrmBF}) can then be canceled by  variations of ${\bf A}$ in the Schwarzian action, ensuring that the total action is differentiable. For this we note from (\ref{lovarnA}) that the latter variations are given by 
\be \delta\int dt S^{(1)}_t (f)=2\int dt \,{\rm tr}\,{\bf  N}\delta{\bf A}\;,\label{dltSch}\ee
${\bf N}$ being the Noether charges  (\ref{bfN}).
  The total action $S_{JT}+ S_{boundary}$ must then also include a  constraint  on the boundary in order that (\ref{btfrmBF}) and (\ref{dltSch}) cancel.  The constraint says that the boundary limit of the scalar field ${\bf B}$ coincides with the Noether charges
\be {\bf B}|_{\partial D}=2{\bf N}\ .\ee
The total action is thus 
\be
S_{JT}+ S_{boundary}=\int_{D} \,{\rm tr}\,{\bf  BF} -\int_{\partial D} dt\,{\cal S}[A]_t(f)+\int_{\partial D} dt\,{\rm tr}\,{\bf\lambda}\left({\bf  B}-2{\bf N}\right)\;,\label{Stotal}
\ee where along with the degrees of freedom ${\bf A}$ and ${\bf B}$, we should include the  boundary field $f$, and here $t$ parametrizes the boundary $\partial D$.  ${\bf \lambda}=\lambda_iT^i$ is a Lie algebra-valued Lagrange multiplier, whose variation imposes the constraint.   Arbitrary variations of ${\bf A}$ in the total action give the boundary terms
\be
\delta S_{JT}+ \delta S_{boundary}=\int_{\partial D} \,{\rm tr}\,{{\bf  B}\delta{\bf  A}} -2\int_{\partial D} dt \,{\rm tr} {\bf N}\delta {\bf A}\;,
\ee
which vanishes due to the constraint, thus insuring differentiability.  

In subsection 4.2, we noted that gauge transformations on  $S^1$ (and after performing a Wick rotation) belong to distinct
homotopy classes associated with the fundamental group $\pi_1\left(SL(2,\mathbb{R})\right)=\mathbb{Z}$. The winding number $n$  is  given by
(\ref{windno}). Gauge field configurations for  ${\bf A}$ with $n=0$ are  topologically trivial.  However for $n$ not equal to  zero  the  gauge field  on $S^1$  cannot be smoothly gauged away by transformations connected to the identity. (More generally, gauge field configurations for  ${\bf A}$ with $n\ne\;$integer cannot be gauged away by topologically trivial or nontrival transformations.)  In the case of JT gravity, the domain of the Schwarzian action is identified with the boundary of the two dimensional domain; so here $\partial D=S^1$.  Extending the nontrivalizable gauge configurations for ${\bf A}$ on $\partial D$ associated with $n\ne $ integer  into the bulk necessarily violates the flatness condition
${\bf F}=0$, and thus a violation of the equations of motion (\ref{wontoo3}), 
somewhere in the interior, 
implying that curvature must be sourced  in the interior.
This provides a natural interpretation of these sectors as corresponding to defects in JT gravity. Such defects are commonly described by insertions that here carry $SL(2,\mathbb{R}) $ holonomy. Examples of such insertions are Wilson lines or punctures that locally violate the flatness of the connection, as discussed in \cite{Mertens:2019tcm},
\cite{Lin:2023wac} and reviewed in \cite{Turiaci:2024cad}. This correspondence indicates that the gauged Schwarzian provides an effective boundary description which is well suited to incorporating these bulk defects.

\section{Concluding remarks}

In this article we obtained a prescription for promoting a nonlinear global symmetry to a gauge symmetry.\footnote{This is in contrast to  work by Balachandran {\it et al.} on nonlinear models which were reformulated as gauge theories.\cite{Balachandran:1978pk}} The fundamental field  transformed with respect to a nonlinear representation of a group. The gauging procedure  was to first construct a composite field ${\bf f}$ from the fundamental field and its derivative, such that it transforms linearly under the {\it global} action of the group.  The next step was to replace ${\bf f}$  by some ${\bf f}_A$ that is covariant under {\it local } transformations, and then finally to apply standard gauging techniques; i.e., replacing ordinary derivatives by covariant ones. Gauge invariant quantities could then be easily constructed.  The procedure was applied  here  to the specific example of the fractional linear representation of $SL(2,\mathbb{R})$, and it gave rise to the gauge invariant analogue of the Schwarzian derivative.   It would be of interest to know the possible relation of our  Schwarzian  derivative extension  to previous ones.\cite{Bouarroudj,Anninos:2021ydw}  Also, since our technique is quite general, it could have application to other nonlinear symmetries.

The composite field ${\bf f}$  proved to be  useful in the study of the Schwarzian dynamics.  It allowed for a simple construction of the  Noether charges associated with the (global) $SL(2,\mathbb{R})$ symmetry of the Schwarzian action.   It may be worthwhile exploring whether analogous composite fields can be constructed to study other nonlinear actions.

The gauge invariant version of the Schwarzian action was constructed with the help of $SL(2,\mathbb{R})$  gauge potentials. These potentials can allow for a straightforward {\it locally} $SL(2,\mathbb{R})$ invariant  coupling to fermionic fields, and an investigation into the  consequences of such a theory could be of interest.  We saw that the potentials could be completely gauged away when the domain is topologically trivial, i.e., $\mathbb{R}$.  This isn't the case however when the domain is $\mathbb{S}^1$.  The latter led to distinct topological sectors, which were labeled by the winding number (\ref{windno}) associated with the fundamental group $\pi(SL(2,\mathbb{R}))=\mathbb{Z}$. When the gauge invariant Schwarzian action determines the
boundary dynamics for two-dimensional gravity, the topological sectors could correspond to  defects in the bulk theory.\cite{Mertens:2019tcm},\cite{Lin:2023wac},\cite{Turiaci:2024cad}   Further investigation of defects from the perspective of the gauged Schwarzian action and its relation to  known work on defects in two-dimensional gravity is warranted.  We hope to address this and related topics in future works.

\bigskip
\bigskip
\appendice{\Large \bf $\;\;$ Alternative expression for the gauge invariant Schwarzian derivative}

\medskip

\noindent
Here we give an alternative expression for the gauge invariant Schwarzian derivative that is equivalent to (\ref{Schw_action_local}).  The result can be written in an expected way in terms of higher order covariant derivatives $(D_A)^n $ of $f$,
\be {\cal S}{[A]}_tf =\frac{ \left(D_A\right)^3 f}{D_A f}-\frac 32 \left(\frac{ \left(D_A\right)^2 f}{D_A  f}\right)^2\label{ginvSch}\ee
The first order covariant derivative of $f$,
  $  D_A f$, was defined in (\ref{CovDer1}).  Because $f$,  and $D_Af$, transforms nonlinearly, care must be taken in defining the higher order covariant derivatives. In order for  (\ref{ginvSch}) to be locally invariant these derivatives should be defined to gauge transform in an analogous manner to ordinary derivatives under global transformations. The specific expressions are
\be\left(D_A\right)^2 f=\ddot f+2\dot f\left( {\rm tr}\,{\bf f}\dot{\bf A} +(A_1+2fA_2)(1+ {\rm tr}\,{\bf fA} )\right)\label{Dt2A} \ee
\beqa \left(D_A\right)^3 f&= &f^{(3)}+3(A_1 +2A_2f)\ddot f +6A_2\dot f^2\left(1 +6  {\rm tr}{\bf fA}  +4 ({\rm tr}{\bf fA})^2\right)\cr&&\cr &&+2\dot f{\rm tr}{\bf f}\ddot {\bf A} +3\dot f(\dot A_1+2  \dot A_2f)+4\dot f(A_1+2A_2f){\rm tr}{\bf f}\dot {\bf A} \cr&&\cr&&+2\dot f \left(\dot A_1+2\dot A_2f+2\,{\rm tr}{\bf A}^2 \right){\rm tr}{\bf fA} +6\dot f {\rm tr} {\bf A}^2\;,\eeqa
where
\be   {\rm tr}\,{\bf fA}=\frac 1{2\dot f}(A_0+A_1f+A_2 f^2)\qquad {\rm trA}^2=\frac12(A_1)^2-2A_0A_2 \ee
\bigskip
{\large{\bf Acknowledgment}}

\noindent
A.P. acknowledges the partial support of CNPq under the grant no.312842/2021-0.


\bigskip
\bigskip

\begin{thebibliography}{}

\bibitem{Ovsienko} V.~Ovsienko, S.~ Tabachnikov,   "What Is . . . the Schwarzian Derivative?" , AMS Notices, 56 (1): 34 (2009).

\bibitem{Osgood1998} B.~Osgood, "Old and New on the Schwarzian Derivative, "In: Duren, P., Heinonen, J., Osgood, B., Palka, B. (eds) Quasiconformal Mappings and Analysis. Springer, New York, NY. 
doi.org/10.1007/978-1-4612-0605-7-16.

\bibitem{Witten:1987ty}
E.~Witten,
``Coadjoint Orbits of the Virasoro Group,''
Commun. Math. Phys. \textbf{114}, 1 (1988)
doi:10.1007/BF01218287.


\bibitem{Alekseev:1988ce}
A.~Alekseev and S.~L.~Shatashvili,
``Path Integral Quantization of the Coadjoint Orbits of the Virasoro Group and 2D Gravity,''
Nucl. Phys. B \textbf{323}, 719-733 (1989)
doi:10.1016/0550-3213(89)90130-2.

\bibitem{Alekseev:2022qrq}
A.~Alekseev, O.~Chekeres and D.~R.~Youmans,
``Towards Bosonization of Virasoro Coadjoint Orbits,''
Annales Henri Poincare \textbf{25}, no.1, 5-34 (2024)
doi:10.1007/s00023-023-01294-1.

\bibitem{Chirco:2024qpx}
G.~Chirco, L.~Vacchiano and P.~Vitale,
``A so(2,2) extension of JT gravity via the Virasoro-Kac-Moody semidirect product,''
[arXiv:2410.10768 [hep-th]].

\bibitem{Carlip:2022fwh}
S.~Carlip,
``A Schwarzian on the stretched horizon,''
Gen. Rel. Grav. \textbf{54}, no.6, 53 (2022)
doi:10.1007/s10714-022-02940-5.

\bibitem{Lin:2019kpf}
H.~W.~Lin and L.~Susskind,
``Complexity Geometry and Schwarzian Dynamics,''
JHEP \textbf{01}, 087 (2020)
doi:10.1007/JHEP01(2020)087

\bibitem{PhysRevLett.70.3339}
S.~Sachdev and J.~ Ye,  ``Gapless spin-fluid ground state in a random quantum Heisenberg magnet,''
 Phys. Rev. Lett.,
  \textbf{70}, 21,
  3339--3342 (1993),
   doi:10.1103/PhysRevLett.70.3339.

\bibitem{A. Kitaev} A.~Kitaev,  “A simple model of quantum holography,” Proceedings of the KITP Strings Seminar and Entanglement 2015 Program (Kavli Institute for Theoretical Physics, Santa Barbara, 2015), http://online.kitp.ucsb.edu/online/entangled15/.


\bibitem{Maldacena:2016hyu}
J.~Maldacena and D.~Stanford,
``Remarks on the Sachdev-Ye-Kitaev model,''
Phys. Rev. D \textbf{94}, no.10, 106002 (2016)
doi:10.1103/PhysRevD.94.106002.

\bibitem{T1983}
C. Teitelboim. ``Gravitation and Hamiltonian Structure in Two Space-Time Dimensions," Phys. Lett. B {\bf126} (1983) 41-45.

\bibitem{J1985}
  R. Jackiw, ``Lower Dimensional Gravity," Nucl. Phys. B {\bf252} (1985) 343-356.



\bibitem{Maldacena:2016upp}
J.~Maldacena, D.~Stanford and Z.~Yang,
``Conformal symmetry and its breaking in two dimensional Nearly Anti-de-Sitter space,''
PTEP \textbf{2016}, no.12, 12C104 (2016)
doi:10.1093/ptep/ptw124.

\bibitem{Turiaci:2024cad}
G.~J.~Turiaci,
``Les Houches lectures on two-dimensional gravity and holography,''
[arXiv:2412.09537 [hep-th]].

\bibitem{Bouarroudj}
Bouarroudj,~S. "Conformal Schwarzian derivatives and conformally invariant quantization." Int. Math. Res. Not. \textbf{29 }, 1553 (2002).

\bibitem{Anninos:2021ydw}
D.~Anninos, D.~M.~Hofman and S.~Vitouladitis,
``One-dimensional Quantum Gravity and the Schwarzian theory,''
JHEP \textbf{03}, 121 (2022)
doi:10.1007/JHEP03(2022)121.

\bibitem{Ozer:2025bpb}
H.~T.~{\"O}zer and A.~Filiz,
``On the explicit asymptotic symmetry breaking of $sl(3,\mathbb {R})$ Jackiw{\textendash}Teitelboim gravity,''
Eur. Phys. J. C \textbf{85}, no.5, 563 (2025).

\bibitem{Mertens:2019tcm}
T.~G.~Mertens and G.~J.~Turiaci,
``Defects in Jackiw-Teitelboim Quantum Gravity,''JHEP \textbf{08}, 127 (2019)
doi:10.1007/JHEP08(2019)127.


\bibitem{Lin:2023wac}
G.~Lin and M.~Usatyuk,
``Revisiting the second order formalism of JT gravity,''
[arXiv:2310.16081 [hep-th]].

\bibitem{Zhang:2025kty}
Z.~Zhang and C.~Peng,
``Gauging the complex SYK model,''
[arXiv:2502.18595 [hep-th]].

\bibitem{Collier:2023fwi}
S.~Collier, L.~Eberhardt and M.~Zhang,
``Solving 3d gravity with Virasoro TQFT,''
SciPost Phys. \textbf{15}, no.4, 151 (2023)
doi:10.21468/SciPostPhys.15.4.151.

\bibitem{Bagrets:2016cdf}
D.~Bagrets, A.~Altland and A.~Kamenev,
``Sachdev{\textendash}Ye{\textendash}Kitaev model as Liouville quantum mechanics,''
Nucl. Phys. B \textbf{911}, 191-205 (2016)
doi:10.1016/j.nuclphysb.2016.08.002.


\bibitem{Mertens:2017mtv}
T.~G.~Mertens, G.~J.~Turiaci and H.~L.~Verlinde,
``Solving the Schwarzian via the Conformal Bootstrap,''
JHEP \textbf{08}, 136 (2017)
doi:10.1007/JHEP08(2017)136.


\bibitem{Galajinsky:2018ona}
A.~Galajinsky,
``A variant of Schwarzian mechanics,''
Nucl. Phys. B \textbf{936}, 661-667 (2018)
doi:10.1016/j.nuclphysb.2018.10.004.

\bibitem{deAlfaro:1976vlx}
V.~de Alfaro, S.~Fubini and G.~Furlan,
``Conformal Invariance in Quantum Mechanics,''
Nuovo Cim. A \textbf{34}, 569 (1976)
doi:10.1007/BF02785666.


\bibitem{Mertens:2018fds}
T.~G.~Mertens,
``The Schwarzian theory \textemdash{} origins,''
JHEP \textbf{05}, 036 (2018)
doi:10.1007/JHEP05(2018)036




\bibitem{FUKUYAMA1985259}
T.~ Fukuyama and K.~ Kamimura, ``Gauge theory of two-dimensional gravity,'' Phys. Lett. B, {\bf 160} 259-262 (1985) .

\bibitem{Isler}
K.~Isler,  C. A. Trugenberger, ``Gauge theory of two-dimensional quantum gravity,'' Phys. Rev. Lett. {\bf 63}  834-836 (1989).

\bibitem{CHAMSEDDINE198975}
A.H. Chamseddine and D. Wyler, ``Gauge theory of topological gravity in 1+1 dimensions,''  Phys. Lett. B,  {\bf 228}  75-78 (1989).


\bibitem{Jackiw:1992bw}
R.~Jackiw,
``Gauge theories for gravity on a line,''
Theor. Math. Phys. \textbf{92}, 979-987 (1992).

\bibitem{Celada:2016jdt}
M.~Celada, D.~Gonz\'alez and M.~Montesinos,
``$BF$ gravity,''
Class. Quant. Grav. \textbf{33}, no.21, 213001 (2016).


\bibitem{Pinzul:2024zkl}
A.~Pinzul, A.~Stern and C.~Xu,
``Embedding space approach to Jackiw-Teitelboim gravity,''
Phys. Rev. D \textbf{110}, no.8, 084033 (2024)
doi:10.1103/PhysRevD.110.084033
[arXiv:2406.05593 [hep-th]].


\bibitem{York:1972sj}
J.~W.~York, Jr.,
``Role of conformal three geometry in the dynamics of gravitation,''
Phys. Rev. Lett. \textbf{28}, 1082-1085 (1972)
doi:10.1103/PhysRevLett.28.1082.

\bibitem{Gibbons:1976ue}
G.~W.~Gibbons and S.~W.~Hawking,
``Action Integrals and Partition Functions in Quantum Gravity,''
Phys. Rev. D \textbf{15}, 2752-2756 (1977)
doi:10.1103/PhysRevD.15.2752

\bibitem{Dyer:2008hb}
E.~Dyer and K.~Hinterbichler,
``Boundary Terms, Variational Principles and Higher Derivative Modified Gravity,''
Phys. Rev. D \textbf{79}, 024028 (2009).

\bibitem{Goel:2020yxl}
A.~Goel, L.~V.~Iliesiu, J.~Kruthoff and Z.~Yang,
``Classifying boundary conditions in JT gravity: from energy-branes to $\alpha$-branes,''
JHEP \textbf{04}, 069 (2021).

\bibitem{Balachandran:1978pk}
A.~P.~Balachandran, A.~Stern and C.~G.~Trahern,
``NONLINEAR MODELS AS GAUGE THEORIES,''
Phys. Rev. D \textbf{19}, 2416 (1979)
doi:10.1103/PhysRevD.19.2416).




\end{thebibliography}
\end{document}